\documentclass[pre,showpacs,twocolumn,amsmath,amssymb,floatfix,bibnotes]{revtex4}
\usepackage{here}
\usepackage[dvips]{graphicx}

\newcounter{Fig}

\begin{document}


\title{Scattering of dipole-mode vector solitons: Theory and experiment}

\author{Wieslaw Krolikowski$^1$, Glen McCarthy$^1$,  Yuri S. Kivshar$^2$, Carsten Weilnau$^3$,
Cornelia Denz$^3$, \\ Juan J. Garc\'{\i}a-Ripoll$^4$, and
V\'{\i}ctor M. P\'erez-Garc\'{\i}a$^4$}

\affiliation{$^1$Laser Physics Center, Research School of Physical
Sciences and Engineering, The Australian National University,
Canberra,
ACT 0200, Australia\\
$^2$Nonlinear Physics Group, Research School of Physical Sciences
and Engineering, The Australian National University,
Canberra, ACT 0200, Australia\\
$^3$ Institute of Applied Physics, Westf\"{a}lische Wilhelms
Universit\"{a}t M\"{u}nster, D-48149 M\"{u}nster, Germany \\
$^4$Departamento de Matem\'aticas, Universidad de Castilla-La
Mancha, 13071 Ciudad Real, Spain}

\begin{abstract}
We study, both theoretically and experimentally, the scattering
properties of optical dipole-mode vector solitons - {\em radially
asymmetric composite} self-trapped optical beams.  First, we
analyze the soliton collisions in an isotropic two-component model
with a saturable nonlinearity and demonstrate that in many cases
the scattering dynamics of the dipole-mode solitons allows us to
classify them as {\em ``molecules of light''} -  extremely
robust spatially localized objects which survive a wide range of
interactions and display many properties of composite states with
a rotational degree of freedom. Next, we study the composite
solitons in an anisotropic nonlinear model that describes
photorefractive nonlinearities, and also present a number of
experimental verifications of our analysis.
\end{abstract}


\maketitle

\section{Introduction}

An understanding of the interaction of simple physical objects
leading to the formation of more complex objects is an ultimate
goal of fundamental research in many fields of physics. Recent
progress in generating {\em spatial optical solitons}--- the
self-trapped states of light with particle-like properties--- in
various nonlinear bulk media allows to study the truly
two-dimensional self-trapping of light and different types of
interaction of multi-dimensional solitary waves, including the
formation of more complicated localized states \cite{book}.

Spatial optical solitons have attracted considerable attention as
possible building blocks of all-optical switching devices where
light is used to guide and manipulate light itself \cite{book}.
The robust nature of spatial optical solitons displayed in the
propagation and interaction \cite{science} allows us to draw an
analogy with atomic physics, treating spatial solitons as ``atoms
of light''.  Furthermore, when several light beams generated by
coherent sources are combined to produce vector or composite
solitons, this process can be viewed as the formation of composite
states or ``molecules of light''.

Recently, the existence of robust ``molecules of light'' in the
form of {\em dipole-mode vector solitons} was predicted
theoretically \cite{us} and also verified experimentally
\cite{exper}. The dipole-mode solitons (or `dipoles', for
simplicity) originate from the trapping of a dipole-mode optical
beam by an effective waveguide created by a mutually incoherent
fundamental beam of nearly radial symmetry.  The first observation
of this novel type of optical vector soliton was reported in Ref.
\cite{exper}, where the dipoles were generated using two different
methods: the phase imprinting technique and a symmetry-breaking
instability of a vortex-mode composite soliton, another type of
fundamental (radially symmetric) composite soliton created by the
incoherent coupling of two optical beams \cite{moti,scat}. It is
worth to mention that vector solitons can be also created in
certain processes involving coherent interaction of waves such as
second harmonic generation. In this particular situation, the two
constituent beams forming a soliton molecule are fundamental and
its second harmonic, respectively \cite{clusters2HG}.

The concept of vector solitons as `molecules of light' can be
compared with photonic microcavity structures, micrometer-size
``photonic quantum dots'' that confine photons in such a way that
they act like electrons in an atom \cite{photonic}.  When two of
these ``photonic atoms'' are coupled together, they produce a
``photonic molecule'' whose optical modes bear a strong
resemblance to the electronic states in a diatomic molecule like
hydrogen \cite{photonic2}. The self-trapped states of light we
study here can be viewed as somewhat similar photonic structures
where, the photonic trap and the beam it guides are both made of
light and create {\em self-trapped photonic atoms and molecules}.

In this paper we present a comprehensive study of the scattering
properties of the dipole-mode vector solitons and analyze, in
particular, the interaction between these objects and other
self-trapped structures such as scalar optical solitons and other
dipoles. We describe a number of interesting effects observed in
numerical simulations of such interactions, for both isotropic
saturable and anisotropic nonlocal nonlinear models. These include
the absorption of a soliton by a dipole and the replacement of the
soliton with a dipole component, the field momentum redistribution
that can be viewed as the transformation of a linear momentum into
an angular momentum with the subsequent dipole spiraling, etc.
Additionally, we verify experimentally some of our analytical
predictions by studying the generation and scattering of the
composite spatial solitons in photorefractive nonlinear crystals.
The versatility of the phenomena described here makes dipole-mode
vector solitons of great importance not only because of  the
fundamental interest in nonlinear physics but also because of
potential promising applications in all optical switching and
integrated optics.

The paper is organized as follows. Section II studies the
scattering of scalar solitons and dipole-mode vector solitons in
the framework of an isotropic saturable nonlinear medium. In this
section, we also present the most important examples of the
dipole-soliton interactions. Section III includes the studies of
the formation and interaction of the dipole-mode solitons in an
anisotropic nonlocal nonlinear model which is used for describing
the nonlocal anisotropic nonlinearities of photorefractive media.
In Sec. IV, we summarize the results of our experimental studies
of the interaction of the dipole solitons in photorefractive
nonlinear crystals. Finally, Sec. V concludes the paper.

\section{Soliton scattering in a saturable isotropic medium}

\subsection{Model and solitons}

We consider here the propagation of two light beams interacting
incoherently in a saturable nonlinear medium. In the steady state
regimes in the paraxial approximation, the mutual beam interaction
can be described by a system of two coupled nonlinear
Schr\"odinger (NLS) equations \cite{us,exper,moti,spiral},
\begin{subequations}
\begin{eqnarray}
i \frac{\partial  u}{\partial z} = -\frac{1}{2} \triangle_{\perp} u + F(I) u, \\
i \frac{\partial  v}{\partial z} = -\frac{1}{2} \triangle_{\perp} v + F(I) v,
\end{eqnarray}\label{Model}
\end{subequations}
where $u({\bf r}_{\perp}, z)$ and $v({\bf r}_{\perp}, z)$ are the
dimensionless envelopes of the beams which are self-trapped in the
cross-section plane ${\bf r}_{\perp} = (x,y)$ and propagate along
the direction $z$.  The function $F(I) = I (1 +s I)^{-1}$
characterizes a saturable nonlinearity of the medium, where $s$ is
a dimensionless saturation parameter $(0 < s <1)$ and $I = |u|^2 +
|v|^2$ is the total beam intensity.

We would like to mention that we consider here only the stationary
propagation of light excluding any nonlinearity-mediated temporal
effects. In fact, the experimental observations indicate that this
is a common situation in many nonlinear systems involving free (no
feedback) propagation of optical beams in both fast (such as
atomic) and slow (photorefractive, thermal) nonlinear media.
Typically, the only dynamics which may occur in such cases is
related to switching effects and dies out within the time scale
determined by the temporal response of the medium. However, the
temporal in noninstantaneous media such as photorefractive
crystals may lead to a number of novel transiting effects (see,
e.g., Refs. \cite{temp,temp0,temp1,temp2}).

Equations (\ref{Model}) describe different types of spatially
localized composite solutions.  {\em The dipole-mode vector
soliton} (or ``a molecule of light'') is a stationary state which
is composed of a nodeless beam in the $v$ component and a dipole
beam (or a pair of out-of-phase solitons) in the $u$ component.
Solitons in the $u$ component have opposite phases and thus they
repel each other, but the role of the complimentary beam $v$ is to
stabilize the structure making it robust.  A numerical analysis of
the linearized equations (\ref{Model}) shows no signs of linear
instability of this composite structure \cite{us}, as was also
recently confirmed by means of the asymptotic analytical theory
\cite{yang}. Moreover, it was shown \cite{us} that such robust
dipole-mode vector solitons exist for a wide range of the beam
powers $P_u=\int |u|^2 d {\bf r}_{\perp}$ and $P_v=\int |v|^2 {\bf
r}_{\perp}$.  Since we are interested in showing stability {\em
far from the regime in which one beam is dominant}, all numerical
experiments are performed using as initial conditions stationary
states in which $P_u \simeq P_v$.

We are interested in the dynamics of the dipole soliton under the
action of finite external perturbations introduced by its
collision with other objects.  The word ``finite'' emphasizes the
fact that we can no longer make use of linearized equations and
that we must deal with the full system (\ref{Model}). This fact,
combined with the complex structure of the dipole which lacks
radial symmetry, makes analytical predictions on the dipole
dynamics very difficult. Nevertheless, as will be shown below, one
may extract some general rules on which qualitative predictions
may be based.

The idea is that the dipole can be seen as a bound state of a
soliton beam (in $v$) plus a pair of vortices with  opposite
charges (in $u$) and, therefore, many of the effects observed in
the composite beam collisions and described below can be
understood once the mutual interaction of these simpler objects is
studied.

One of the components of the dipole is a soliton beam (to be
referred to as {\em soliton} hereafter). Spatial solitons are
stable localized states which have no nodes and which are the
states of minimum energy of the system {\em for a fixed power}.
When two of these solitons are in different, mutually incoherent
beams (say,  one in $u$ and the other one,  in $v$), they interact
incoherently and {\em attract each other}. Thus, during an
incoherent interaction two solitons may either become bound or
scatter. In the former case, we have an example of what we call a
{\em molecule of light}, which is typically referred to as a
``vector soliton''. However, when two solitons are derived from
the same beam they interact coherently and the outcome of  their
mutual interaction depends  on their phase difference.  When this
quantity is small or zero,  solitons  experience mutual {\em
attraction}, whereas if their mutual phases differ by $\pi$, they
{\em repel each other}.

Another nonlinear structure that should be mentioned in this
context is {\em a vortex-mode composite soliton} introduced in
\cite{moti} which in our model (\ref{Model})  is stable only in
the vicinity of the bifurcation point \cite{yang}. Thus, the
vortices may only be stabilized by co-propagating  with a very large
soliton beam (e.g., when a vortex in the linear beam $u$ is guided
by an effective waveguide created in the component $v$).
Otherwise, a composite state of a vortex plus a soliton constitutes
an unstable molecule of light.

A dipole can be seen as a pair of vortices as described above or,
alternatively, as  a bound state of two solitons with a phase
difference of $\pi$. While in principle, these solitons should
repel each other, the system is stabilized due to the interaction
with a soliton-induced waveguide created by the other, mutually
incoherent, component.

\subsection{Numerical results for the soliton collisions}

\subsubsection{Soliton-dipole scattering}

The first type of numerical simulations we present here consists
in shooting a scalar soliton against a dipole-mode vector soliton.
All the simulations discussed here have been performed using a
split-step operator technique using FFT, with grid sizes of up to
512$\times$512 points covering a rectangular domain of
68$\times$34 adimensional units.  The initial data are always a
combination of stationary states. For instance, when a soliton is
launched against a dipole, we start with
\begin{mathletters}
\begin{eqnarray}
u({\bf x},0) & = & u_{dipole}({\bf x}) + u_{soliton}({\bf x}-{\bf d})
e^{-i{\bf p_0 x}},\\
v({\bf x},0) & = & v_{dipole}({\bf x}).
\end{eqnarray}
\end{mathletters}
Here ${\bf d}=(d_x,d_y)$, $d_x \gg d_y$, $d_y$ is {\em the impact
parameter},  and $\mathbf{p_0}$ is proportional to the initial
(linear) momentum of the incoming scalar soliton. The initial data
$u_{dipole}$, $u_{soliton}$,  and $v_{dipole}$ are obtained
numerically by a suitable minimization procedure outlined in Ref.
\cite{us}.


The result is an inelastic collision in which the soliton becomes deflected and
the dipole gains both {\em linear} and {\em angular} momenta. The whole process
is depicted in Fig.1. Soliton scattering occurs when the incident
beam has medium to large linear momentum or when it has an appropriate initial
phase. For instance, in Fig.1 the incident soliton has sign $(-)$
and it crashes against the part of the dipole with $(+)$ sign. A conservation
law forces the dipole to rotate and the soliton becomes deflected, sometimes as
much as by a 90 degree angle.


When the linear momentum of the incident soliton is large, it
moves too fast to suffer a destructive influence from the dipole.
In Fig.2 we plot the exchange of the linear momentum
between the soliton and dipole as a function of the impact
parameter. The effective interaction is clearly {\em attractive}:
the soliton coming from below ($d_y < 0$) feels the drag of the
dipole above it and gets deflected upwards ($p_y > 0$), while the
dipole moves downwards.


The second family of numerical experiments is performed with
solitons which are slow and, as is usual in scattering processes,
the effects of the interaction process may be more drastic. For
some impact parameters the soliton gets too close to the lobe of
the dipole with the smallest phase difference and fuses with it
with some emission of radiation and a subsequent rotation of the
dipole. This is well reflected in Fig.3 (radiation
is not seen).

\subsubsection{Dipole-dipole collisions}

The third family of numerical simulations corresponds to shooting
dipoles against each other. These collisions, which resemble interaction of atomic molecules provide a rich source
of phenomena depending on the mutual orientation of the dipoles
and on the initial energy. Figure 4 summarizes the main
results observed.  There we see {\em three cases} (a-c) in which
the dipole solitons are preserved. The figure shows an in-phase
collision with weak interaction [Fig.4(a)], an
out-of-phase collision with repulsion [Fig.4(b)], and
an example of the collision with nonzero impact parameter in which
two vortex states are created and they decay into a pair of
spiralling solitons [Fig.4(c)].

The last case, Fig.4 (d), shows an interesting
inelastic process when two dipoles fuse into a more complex state
which then decays creating a new dipole and a pair of simple
solitons.  All these processes may be understood in terms of the
phase of the lobes of each dipole as described above.

\section{Soliton scattering in anisotropic nonlocal media}

\subsection{Composite solitons}

The dipole-mode vector solitons considered so far were restricted
to those realized in isotropic nonlinear media.  However, up to
now the majority of experimental observations of dipole-mode and
multi-pole vector solitons have been performed in photorefractive
nonlinear crystals which are known to exhibit  anisotropy in their
nonlinear response \cite{anisotropy}. In effect, even circularly
symmetric optical beams induce strongly asymmetric refractive
index changes which significantly affect the formation of spatial
solitons as well as their interaction.

In this section we employ the commonly accepted model for the
photorefractive nonlinearity that takes into account its most
important properties   \cite{wieslaw} to investigate some of the
previously discussed examples of interactions of vector solitons.

The interactions we consider here involve dipole-mode vector
solitons and scalar solitons. The dipole-mode vector solitons
consist of two mutually incoherent optical beams with the
envelopes $u$ and $v$, propagating in a bulk anisotropic nonlocal
nonlinear medium such as a biased photorefractive crystal. When
the characteristic spatial scales are larger than the
photorefractive Debye length and the diffusion field may be
neglected, the steady-state propagation along the $z$ axis of a
photorefractive crystal with an externally applied electric field
along the $x$ axis is described by the equations:

\begin{equation}
\label{eq1}
  \begin{array}{l}
      {\displaystyle
        i\frac{\partial u}{\partial z} + \frac{1}{2}\nabla^2 u=
          -\frac{\gamma}{2}\frac{\partial\varphi}{\partial x}u,} \vspace{3mm}
 \\
      {\displaystyle
     i\frac{\partial v}{\partial z} + \frac{1}{2}\nabla^2 v=
          -\frac{\gamma}{2}\frac{\partial\varphi}{\partial x}v,}  \\
      {\displaystyle
        \nabla^2\varphi+\nabla\varphi\nabla\ln (1+I)=
          E_0x_0\frac{\partial}{\partial x}\ln(1+I),}
  \end{array}
\end{equation}
where $\gamma$ and $E_0$ are the normalized nonlinearity coefficient and
external field, respectively, $I\equiv |u|^2+|v|^2$ is the total
intensity,
$\nabla ={\hat x}(\partial/\partial x)+{\hat y}(\partial/\partial y)$,
and $\varphi$ is the dimensionless electrostatic potential induced by the
light with the boundary condition
$\nabla \varphi(\vec r\rightarrow\infty)\rightarrow 0$.
The propagation coordinate $z$ is measured in units of the diffraction
length, and the transverse coordinates are normalized by the
characteristic beam size $x_0$. The above system of equations was solved numerically  by applying concurrently the finite difference and  split-step Fast Fourier methods to the electrostatic potential and  propagation equations, respectively and using a  256x256 grid size computational window.

\subsection{Soliton-dipole scattering}

The first two of the numerical simulations we present here consist
of colliding a scalar soliton into one lobe of a dipole-mode
vector soliton.  The scalar soliton is coherent with the dipole
component of the vector soliton and can thus be considered part of
the same beam in the theoretical model (\ref{eq1}) above. The
initial data for all cases presented in this section reflect the
experimental conditions in that they are not exact solutions to
the model. The fundamental component of the vector soliton and the
scalar beam are Gaussian beams and the dipole component is the
first derivative of a Gaussian beam along $y$ coordinate.


The first example of the collision between the dipole-mode vector
soliton and the  scalar soliton is shown in Fig.5.
Here the scalar soliton is in phase with the lobe of the dipole with which it
collides. This leads to a strong attraction between  the scalar beam and
the lobe and eventual absorption of the former followed by, as in
the isotropic case, a rotation of the whole dipole-mode vector
soliton. In a photorefractive crystal the complete rotation of a
dipole  soliton is prohibited by the anisotropy of the nonlinear
refractive index change \cite{optlet}. Hence, unlike the isotropic
case (see Fig.1), the vector soliton exhibits only
angular oscillations about the vertical axis.  As
Fig.5 clearly shows the collision leads to dipole
soliton experiencing  also a lateral shift which is due to
transfer of a linear momentum from the scalar soliton to the
vector soliton. As our numerical simulations show the outcome of
this collision can be  more dramatic if the intensity of the
scalar beam is increased. In such a case the collision may result in the
break up of the vector soliton such that the two out-of-phase
lobes of the dipole are no longer trapped. Such  an  effect occurs
when the intensity of the scalar beam is  comparable to that of
the fundamental component of the vector soliton.
%

In the next example shown in Fig.6 the scalar
soliton and the lobe of the dipole it interacts with are
out-of-phase. Both solitons propagate initially along parallel
trajectories. Because of the phase relation the interaction is now
repulsive leading to rotation of the dipole. Again the rotation is
limited by the anisotropic refractive index distribution. On
further propagation, the vector and scalar solitons are clearly
repelled and the vector soliton will reorientate to the stable
vertical direction once the effect due to presence of the scalar
beam is negligible. Our simulations show that by increasing the
angle between initial trajectories of both solitons one can induce
even stronger rotation of the dipole but this constitutes so
drastic perturbation to the dipole that it often breaks-up so that
the dipole lobes are no longer trapped.

\subsection{Dipole-dipole scattering}

The next few examples  of the numerical simulations involve the
collisions of two dipole-mode vector solitons.  Unless stated otherwise,
 the dipole components of the vector solitons are
coherent with each other while the  fundamental components are
mutually incoherent.


In the example depicted in Fig.7 the two
identically oriented vector solitons  collide centrally
propagating along the direction of the applied electric field. The
mutual interaction is now  attractive (phases of dipoles coincide)
and since the  intersection angle (or transverse velocities) of
the solitons is relatively small the solitons fuse forming a new
dipole-mode vector soliton. This fused soliton undergoes internal
oscillations as its parameters are quite different from the exact
soliton solution. Further simulations show that after  emitting
some radiation  the structure reaches a steady state.


In Fig.8 the phase of one of the dipoles has been
shifted by $\pi$ and hence each lobe now collides with an out of
phase lobe of the other dipole. This  leads to the mutual
repulsion of the dipole-mode vector solitons, which the weak
attraction of the incoherent Gaussian beams is unable to
counteract. Further simulations show that the structure of each
soliton is well preserved throughout the interaction and therefore
this collision could be considered as an almost elastic collision.
Note very close similarity between this result and simulations of dipoles interaction in isotropic  medium shown in Fig.4.

If the initial trajectories of both solitons are chosen such that
the collision is non-central   then both dipoles usually undergo
spatial rotation as shown in Fig.9. This rotation
is initiated by the mutual repulsion of the out-of-phase lobes in
respective dipole components.  Again, the rotation is limited by
the anisotropic induced refractive index change as the dipoles
force their way past each other. As the solitons pass the zone of
interaction they undergo oscillatory rotation about the vertical
axis which is again due to the orientation perturbation caused by
the collision as can be seen in (d),(e) and (f).


Fig.10 shows another example of the
noncentral collision of the two dipoles for the case when the
initial angle between soliton trajectories is increased to 0.21
degree. As the impact parameter (i.e. lateral mismatch of initial
trajectories) is smaller than in the preceding case, the repulsive
interaction  of dipole components is weaker. This, combined with the
larger mutual velocity of the solitons (larger intersection angle)
allows the solitons to pass through one another with a weak, but still
visible, intermediate tilt of their axes. The final state
corresponds to the perturbed dipole states with an excited internal
oscillation which is reflected in an unequal energy distribution
between both lobes in each dipole.

Finally, Fig.11 shows an example of two dipole-mode
vector solitons colliding along their long (y) axes. In this
particular example  the  fundamental components of both dipole
solitons are coherent and in-phase. The initial orientation of the
dipoles was such that the two directly interacting lobes were
out-of-phase hence their interaction was repulsive. On the other
hand  the interaction of fundamental components  is strongly
attractive. Since the mutual velocity of the solitons is less
(smaller intersection angle) the fundamental components  fuse in
collision  trapping two out-of-phase lobes. In this way new dipole
mode vector solitons was formed with its dipole component being
constituted by two out-of-phase lobes coming from two different
solitons. Notice that because of symmetry of the problem  this new
vector soliton does not exhibit any transversal motion.  The
remaining the outmost  lobes are  no longer bound and separate as
fundamental solitary beams as can be seen in
Fig.11(e).


\section{Experimental results}

The generation of an isolated dipole-mode vector soliton was reported
earlier in Ref. \cite{exper,anton,motzek}.
Fig.12  shows schematically experimental set-up
used in those and current studies.

 The dipole-mode
soliton can be created using a few different processes: (i) {\em
phase imprinting}, when one of the beam components is sent through
a phase mask in order to imprint the required phase structure, and
(ii) {\em symmetry-breaking instability} of a vortex-mode
composite soliton or (iii) superposition of two oppositely charged
vortex beams.  In this way, one obtains a dipole-like structure
with a phase jump along its transverse direction that is
perpendicular to the optical axis of the crystal [see Fig.13(a)].  This dipole-like beam is then combined with
the second, node-less beam and the resulting composite beam is
focused on the input face of the photorefractive SBN crystal (the
crystal has the same parameters as in \cite{exper}), biased with a
DC field of 1-2.5 kV applied along its optical axis. The exit and
input facets of the crystal are viewed by CCD cameras and images
stored in the computer. To control the degree of saturation of the
nonlinearity, we illuminated the crystal with a wide beam derived
from a white light source. In our experiments the ratio of the
peak intensity of the  soliton beam to intensity of white light
illumination was always of the order of unity. To ensure that both
beams forming a composite structure are mutually incoherent, one
of them is reflected by a mirror mounted on a piezo-electric
transducer (PZT). When the transducer is driven by a 1kHz signal
it introduces a phase modulation   into the beam. As the
photorefractive crystal is slow, it cannot follow fast changes of
the relative phase of both beams making them effectively
incoherent. Propagating in a {\em self-focusing saturable medium},
such a composite input beam creates a dipole-mode vector soliton,
as shown in Fig.13(b-d).   As discussed above, the
fundamental component creates an effective asymmetric waveguide
that guides a dipole-like mode in the form of two out-of-phase
solitary beams that otherwise would repel and fly apart.


To observe the soliton-dipole interaction effects, we launched a
scalar soliton beam against the dipole soliton. The input state is
shown in Fig.13(e), where the dipole-mode soliton is
presented by its two-lobe $u$ component only. The relative phase
of the soliton and the upper lobe of the dipole is close to $\pi$.
Therefore, when the soliton interacts with a dipole, it gets
deflected (repulsive interaction) and transforms a part of its
linear momentum into an angular momentum of the dipole that starts
rotating clockwise which is clearly visible in Fig.13 (f-h). 
When the scalar soliton is removed, the
vector soliton  rotates back until it realigns to its original
stable orientation orientation which is  vertical in the
experimental situation discussed here. Fig.14
illustrates the dynamics of this process showing the intensity of
both, dipole (top row) and fundamental (bottom row) components. It
should be noted that the above experimental observations are in
accordance with the results of numerical modelling. In particular,
the intensity distributions shown in  Fig.13(e-h)
closely resemble theoretical results displayed  in
Fig.1(a) and Fig.6(c) Also, the
observed dynamical behavior of the dipole agrees well with our
(not shown here) numerical simulations.



In the subsequent graph (Fig.15) we show the
interaction of two closely spaced  dipole mode vector solitons.
These solitons were propagating in parallel. Graphs
Fig.15(a-b) shows the intensity of dipole and
fundamental components at the input face of the crystal.
Fig.15(c) depicts the dipole components of both
solitons seen at the exit of the crystal. This picture was
obtained by superimposing two graphs corresponding to different
solitons, hence  it  displays propagation {\em without}
interaction. The dipole-soliton interaction is shown in the bottom
row of this figure. Graphs in Fig.15(d-f) show the
total  intensity (d) and the intensity of dipole (e) and
fundamental components of the interacting dipole-mode solitons.
Because of $\pi$ phase difference between nearest lobes the
interaction leads to a repulsion of the corresponding lobes and
the clockwise rotation of both vector solitons. This behavior is
analogous to that found in our numerical simulations shown in
Fig.9(b-c). In both, theory and experiment, the
mutual rotation of the interacting dipoles is evident.

\section{Conclusions}

We have studied interactions of dipole-mode composite solitons
with different nonlinear localized structures such as scalar
solitons and other dipole-mode solitons. Our studies demonstrate
that, apart from the robustness of the dipole-mode solitons
against strong perturbations such as the interaction with other
solitons (which is a consequence of their linear stability
predicted earlier), in many cases the dynamics of the dipole-mode
composite solitons can be understood qualitatively as that of the
bound states of simpler solitons. Such dynamics involves two major
degrees of freedom of the composite soliton, namely, the rotation
of the structure as a whole and the relative oscillation of the
lobes of the dipole, which both can be excited in collisions. In
some cases, the excitation of the dipole soliton is so strong that
the dynamics of the composite structure resembles that of a pair
of spiraling beams analyzed earlier in Ref. \cite{spiral}. This is
only one of many interesting phenomena observed in our
numerical simulations of dipole-mode soliton scattering which
also include the excitation of rotational motion by collision with
a scalar soliton, annihilation or strong deflection of the
incident soliton, etc. Even richer effects are observed when two
dipole-mode solitons collide with a nonzero impact parameter. It
is remarkable that the rich dynamics observed here may be
understood qualitatively in terms of the structure of the
colliding objects and the relative phases of the dipole
components. To make our results more realistic providing a
comparison with experiment, we have extended our analysis and have
studied the anisotropic nonlocal model that is more relevant for
describing photorefractive nonlinearities.  Finally, we have
verified  some of our theoretical predictions experimentally
employing the self-trapping effect in nonlinear photorefractive
crystals.

A rich variety of the scattering effects described theoretically
and verified experimentally might make the dipole-mode solitons
attractive candidates for realistic applications in the field of
integrated optics.

\section*{Acknowledgements}

This work was partially supported by the Planning and Performance
Fund of the Australian National University, the Australian
Research Council, the grant BFM2000-0521 (Ministerio de Ciencia y
TecnologÌa) and the grant PAC02-002 (Consejeria de Ciencia y
Tecnologia de la Junta de Comunidades de Castilla-La Mancha),
Australian-German Joint Reserach Co-operation Scheme,
Australian-German project exchange program by DAAD, and the
DFG-Graduiertenkolleg "Nichtlineare kontinuierliche Systeme".

\newpage
{\large Figure Captions}
\begin{itemize}
\item
Fig.1. Soliton-dipole scattering. (a) Snapshots of the intensity profile of each
  of the beams taken at few points along  their propagation distance.  (b) Three-dimensional plot of the total intensity $|u|^2+|v|^2$,
  which shows the dipole rotation induced in the collision. (c) Same as (b), but with  the $u$ and $v$ components shown separately.

\item
Fig.2 (a, b) Components of the linear momentum of the incident soliton (solid line) and dipole (marked by circles) after an inelastic collision with a large
  incident momentum, $\mathbf{p_u} \equiv \int u^* \nabla u \, d{\bf
    r}_{\perp}$, as a function of the impact parameter $d_y$, which shows the
  crucial role of the dipole asymmetry. Total $p_y$ does not vanish because of
 radiation (not seen in the figure).

\item
Fig.3 Absorption of a soliton by a dipole. (a) Intensity profile of
  each of the beams at various value of their propagation distance --the darker the more intense; (b) Three-dimensional  plot of the total beam intensity; (c) Same as in (b) with the $u$ and $v$ components separated.

\item
Fig.4 Collisions of two dipoles with the zero [(a), (b), and (d)] and nonzero [(c)] impact parameter, and different orientation of the dipole prior the
  collision [cf. (a) and (d)].

\item
Fig.5 Soliton absorption in the anisotropic nonlocal model. The
top row shows the dipole components of the vector solitons while
the bottom row shows the Gaussian components when a colliding
soliton is absorbed by the dipole. Relative intensities:
Gaussian=1.4, Dipole=1.1, Coherent beam=1.0. Frames are taken at
$z=$0.0, 1.1, 2.6, 4.1, 5.6 diffraction lengths, respectively.
Collision angle=$0.24^o$. The applied field is in the horizontal
direction.

\item
Fig.6 Soliton-dipole interaction in the anisotropic model.  The
top row shows (a) initial dipole components, (b) stable
orientation of the dipole components with no interacting scalar
soliton at $z=$2.3, and (c) rotated dipole when in the presence of
a scalar soliton also at $z=$2.3. The scalar soliton is of out of
phase with the dipole lobe closest to it. The bottom row shows the
corresponding fundamental components. The applied field is in the
horizontal direction. Relative intensities: fundamental component
=1.2, dipole=1.1, scalar beam=1.0.

\item
Fig.7 Dipole fusion in the anisotropic model. The top row shows
the dipole components of the vector solitons while the bottom row
shows the Gaussian components. Relative intensities:
fundamental=1.0, dipoles=1.8. Frames are taken at $z=$0.0, 3.4,
4.8, 5.9, 7.0, respectively. Collision angle=$0.13^o$.

\item
Fig.8 Elastic collision of two dipoles in the anisotropic
model. The top row shows the dipole components of the vector
solitons, while the bottom row shows the Gaussian components.
Relative intensities: Gaussian=1.0, Dipole=1.8. Frames are taken
at $z=$0.0, 1.5, 3.0, 4.5, 6.0, respectively. Collision
angle=$0.21^o$.

\item
Fig.9 Dipole rotation in the anisotropic model.  The top row
shows the dipole components of the vector solitons while the
bottom row shows the Gaussian components. Relative peak intensities:
Gaussian=1.0, Dipole=1.2. Frames are taken at $z=$0.0, 2.3, 3.4,
4.5, 5.6, 6.7, respectively. Collision angle=$0.19^o$.

\item
Fig.10 The same as in Fig.9 but for the
collision angle =$0.21^o$. Frames are taken at $z=$0.0, 2.3, 3.0,
3.7, 4.4, respectively.

\item
Fig.11 Dipole collision in the anisotropic model. The top row
shows the dipole components of the vector solitons while the
bottom row shows the fundamental components. Relative intensities:
Gaussian=0.65, Dipole=1.1. Frames are taken at $z=$0.0, 1.5, 3.7,
5.2, 6.7, respectively. Collision angle=$0.17^o$.

\item
Fig.12 Experimental setup. PM-phase mask, F-filter, $\lambda/2$
- half-waveplate, P-polarizer, PBS-polarizing beam splitter,
PZTM-mirror mounted on piezo-electric transducer, O- microscope
objective, F-filter, V-DC biasing field applied to the crystal,
CCD - camera.

\item
Fig.13 Experimental results. Top row: the formation of the dipole
soliton.  (a) Initial intensity of the dipole component; (b) total
intensity of the vector soliton after 10mm propagation in a biased
SBN crystal; (c,d) the dipole and fundamental components of the vector
soliton after propagation. Bottom row: interaction of dipole and
scalar solitons. (e) Initial intensity of the dipole and soliton
beams; (f) the same after interaction in a biased SBN crystal;
(g-h) dipole and fundamental components of the rotated vector
soliton after the interaction. Voltage: 1.3 kV. Intensities:
dipole- 1.5$\mu W$, the fundamental component - 0.9 $\mu W$,
scalar soliton - 1.6$\mu W$.

\item
Fig.14 Temporal evolution of the dipole component of the tilted
vector soliton after the scalar soliton is blocked.  (a) initial
intensity of the scalar soliton and vector solitons; (b) dipole
component immediately after the scalar beam is blocked; (c-g)
temporal evolution of the dipole component. The time step between
frames is 0.5sec.

\item
Fig.15 Experimental results for the interaction of two dipoles.
Top row: (a-b) intensity distribution of the two dipoles and two
fundamental components; (c)superimposed images of the dipole
components of two independently propagating vector solitons (no
interaction); Bottom row: soliton interaction. (d) total intensity
distribution of two interacting solitons; (e-f) their dipole and
fundamental components. Voltage: V=1.1kV. Intensities of both
dipoles: 1.5, 1.6 $\mu W$ Intensities of the fundamental
components: 1.6, 1.7$\mu W$.
\end{itemize}
\end{document}